# Are They All Good? Studying Practitioners' Expectations on the Readability of Log Messages

Zhenhao Li
Concordia University
Montreal, Canada
l_zhenha@encs.concordia.ca

An Ran Chen
University of Alberta
Edmonton, Canada
anran6@ualberta.ca

Xing Hu*
Zhejiang University
Ningbo, China
xinghu@zju.edu.cn

Xin Xia
Zhejiang University
Hangzhou, China
xin.xia@acm.org

Tse-Hsun (Peter) Chen
Concordia University
Montreal, Canada
peterc@encs.concordia.ca

Weiyi Shang
University of Waterloo
Waterloo, Canada
wshang@uwaterloo.ca

*Abstract*—Developers write logging statements to generate logs that provide run-time information for various tasks. The readability of log messages in the logging statements (i.e., the descriptive text) is rather crucial to the value of the generated logs. Immature log messages may slow down or even obstruct the process of log analysis. Despite the importance of log messages, there is still a lack of standards on what constitutes good readability of log messages and how to write them. In this paper, we conduct a series of interviews with 17 industrial practitioners to investigate their expectations on the readability of log messages. Through the interviews, we derive three aspects related to the readability of log messages, including *Structure*, *Information*, and *Wording*, along with several specific practices to improve each aspect. We validate our findings through a series of online questionnaire surveys and receive positive feedback from the participants. We then manually investigate the readability of log messages in large-scale open source systems and find that a large portion (38.1%) of the log messages have inadequate readability. Motivated by such observation, we further explore the potential of automatically classifying the readability of log messages using deep learning and machine learning models. We find that both deep learning and machine learning models can effectively classify the readability of log messages with a balanced accuracy above 80.0% on average. Our study provides comprehensive guidelines for composing log messages to further improve practitioners' logging practices.

## I. INTRODUCTION

Software logs are important source of information in software systems that record system run-time behaviors. Developers can leverage the valuable information in logs to assist in many tasks, such as program comprehension [1], [2], [3], anomaly detection [4], [5], [6], and failure diagnosis [7], [8], [9], [10], [11], [12]. Logs are generated from logging statements inserted by the developers. For example, in a logging statement from Elasticsearch [13]: *logger.info("Successfully updated remote job [{}]", update.getJobId());* the logging statement is written in Java using Log4j [14] framework, the verbosity level is *Info*, the log message is *"Successfully updated remote job [{}]"*, and the dynamic variable is the value of `update.getJobId()`.

* Corresponding author.

The value of logs highly relies on the quality of log messages (i.e., the part of *"Successfully updated remote job [{}]"* in the example above). Developers leverage the information in log messages as clues for debugging and failure diagnosis, unclear log messages may confuse developers and further slow down or even obstruct the process of log analysis [15]. For example, if the log message is only *"Shutting down."*, it is still difficult to know what is shutting down.

Prior studies provide some supports on composing logging statements, e.g., where to insert logging statements [16], [17], [18], [19], [20], how to choose the verbosity level [21], [22], [23], and generating logging statements by learning from existing data [24], [25]. However, to the best of our knowledge, there is still a lack of practical standards or systematical investigation on what are "good" log messages that record valuable information and are easy to comprehend. Therefore, how to compose log messages with good readability that can clearly and sufficiently record system run-time behaviors is still an on-going challenge. The reliability of automated recommendations learned from log messages with inadequate readability might also be decreased.

In this paper, we conduct a comprehensive study to investigate practitioners' expectations on the readability of log messages and seek possible improvements: 1) We first conduct a series of semi-structured interviews with 17 industrial practitioners from 11 companies worldwide to gain insights on their perspectives of log messages' readability; 2) We manually study the readability of log messages in nine large-scale open source software systems; 3) We validate our findings from the interviews and manual studies through an online questionnaire survey with 56 participants; 4) We further explore the potential of automatically classifying the readability of log messages using deep learning and machine learning approaches.

In particular, we study the following three research questions:

**RQ1: What are practitioners' expectations on the readability of log messages and how to improve it?** By analyzing the interview records, we derive three aspects that are related to the readability of log messages, including ***Structure***, ***Information***, and ***Wording***. For each aspect, we also derive several specific practices that can be used to improve the readability. Our survey participants acknowledge the importance of these aspects and the effectiveness of improvement practices. Among the three aspects, *Information* is considered as the most important aspect: 87.5% of the participants consider it is "Very important" and 12.5% consider it is "Important".

**RQ2: How is the readability of log messages in large-scale open source software systems?** We use the data set of logging statements provided by a prior study [22] to manually investigate the readability of log messages based on the three aspects discussed in RQ1. We find that only 61.9% of the log messages on average have adequate readability in all three aspects, meaning that a large portion of the log messages (i.e., 38.1%) in these systems have inadequacy in terms of their readability, e.g., 21.7% of the log messages are inadequate in the aspect of *Information*.

**RQ3: Can we automatically classify the readability of log messages?** We explore the potential of automatically classifying whether a log message has readability issue or not using several deep learning and machine learning approaches (e.g., Bi-LSTM, Random Forest, and Decision Tree). We find that both deep learning and machine learning approaches can effectively classify the readability of log messages (e.g., Bi-LSTM and Random Forest achieve a balanced accuracy of 82.1% and 86.3% on average, respectively).

The contributions of this paper are as follows:

- We are the first study that investigates the readability of log messages by conducting interviews with industrial practitioners. We derive three aspects that are related to the readability of log messages and several corresponding practices to improve the readability for each aspect.
- We find that a large portion of the log messages in large-scale open source systems actually have inadequate readability. Future works should consider this issue when leveraging existing data for automated recommendation or generation.
- We explore the potential of automatically classifying the log messages whose readability might need improvement and achieve encouraging results.

Overall, our study provides a systematic comprehension on the readability of log messages and sheds light for future studies on uncovering empirically-derived standards to guide developers' logging practices.

**Paper Organization.** Section II summarizes the related work. Section III describes the research methodology of our study. Section IV presents the results by answering three research questions. Section V discusses the implications of our study. Section VI discusses the threats to validity of our study. Section VII concludes the paper.

## II. RELATED WORK

**Empirical Studies on Logging Practices.** Yuan et al. [26] studied the logging practices in C/C++ applications and found that developers often improve log messages as after-thoughts. Chen et al. [27] further studied the logging practices in Java applications and pointed out the similarities and differences of logging practices compared to C/C++ applications. Some prior studies also empirically studied the logging practice in Android applications [28], Linux kernel [29], and test code [30]. Other prior studies focused on assisting developers in making logging decisions and improving the logging practices [31], [32]. For example, Fu et al. [17] and Li et al. [18] investigated where logging statements were placed to identify the common categories of logging locations. Zhu et al. [16] proposed an automated tool for suggesting logging locations. Several prior studies [21], [22], [23] proposed automated approaches to help developers select the appropriate verbosity level. Li et al. detected the logging code smells related to duplicate logging statements [33] and studied their relationships with code clones [34]. These studies focus on empirically studying logging practices or provide supports for deciding the logging locations or verbosity levels. In our study, we investigate practitioners' expectations on the readability of log messages, which complement prior studies on improving logging practices.

**Studies on Log Messages in Logging Statements.** He et al. [35] empirically studied the n-gram patterns of log messages and proposed an information retrieval based approach that generates log messages from similar code snippets. Ding et al. [25] formed the process of log message generation as neutral machine translations and achieved promising results in such generations. Mastropaolo et al. [24] proposed a deep learning based approach that can generate complete logging statements, including log messages for Java methods. Despite the extensive studies on log messages in logging statements, the readability of those log messages has not been investigated thoroughly. In this paper, we systematically study the readability of log messages and derive three aspects related to the readability. For each aspect, we also derive several improvement practices based on our interviews with industrial practitioners.

## III. RESEARCH METHODOLOGY

As shown in Figure 1, our research methodology consists of four stages. *Stage 1:* Semi-structured interviews [36], [37] with practitioners from industry on their experiences in reading log messages, their perspective on the readability of log messages and how to improve it. *Stage 2:* Manually study how prevalent are log messages that may need improvement based on the aspects of readability derived from the interview results. *Stage 3:* A questionnaire survey [38], [39] for confirming the aspects of log message's readability with the corresponding improvement practices that are summarized from the interview, and verifying the manual investigation results in the prior stage. *Stage 4:* Exploring the potential of automatically classifying the readability of log messages.

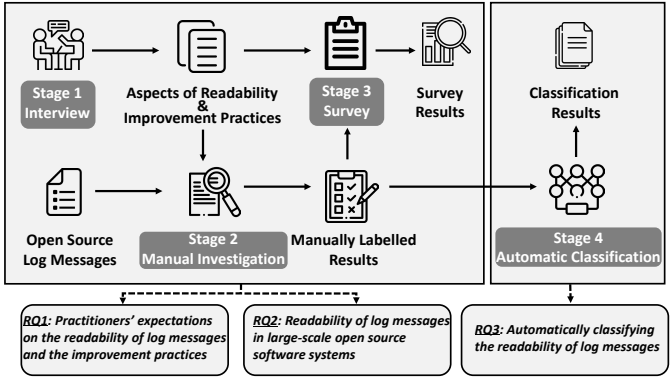

Fig. 1. Overview of our study.

TABLE I
AN OVERVIEW OF THE DATA SET. LOC: LINES OF CODE, NOL: NUMBER OF LOGGING STATEMENTS.

| System | LOC | NOL | Sample |
|---|---|---|---|
| Cassandra | 432K | 1,316 | 298 |
| ElasticSearch | 1.50M | 2,619 | 337 |
| Flink | 177K | 2,455 | 333 |
| HBase | 1.26M | 5,524 | 360 |
| JMeter | 143K | 1,848 | 319 |
| Kafka | 267K | 1,563 | 308 |
| Karaf | 133K | 706 | 251 |
| Wicket | 216K | 413 | 201 |
| Zookeeper | 97K | 1,245 | 295 |
| Total | 4.2M | 17,689 | 2,702 |

## A. Stage 1: Interview

In our interview with industrial practitioners, we investigate their perspective on the readability of log messages and their expectations on the specific practices that can improve the readability.

**Interview Process.** We first develop an interview guideline by gathering all the authors of this paper and brainstorming a set of open-ended questions. All of the authors have industrial experience and are proficient in the knowledge of logging. The first author of this paper then follows the guideline and conducts a series of individual interviews using online video-conferencing tools with 17 software practitioners. Before the start of each interview, we first send the introduction part of the guideline to the interviewees to let them know the background information of our study, ensure that they are aware of the interview being recorded, and emphasize that we will protect the participants' identities. Each interview takes 30-40 minutes and is semi-structured with three parts of questions[1].

*Part 1:* We ask some questions about the interviewees' background information (e.g., years of experiences, role of responsibility, and programming languages used in daily job).

*Part 2:* We ask open-ended questions about their experiences in reading and analyzing log messages (e.g., *"What kind of information provided by the log messages is important to you?"*, *"Have you ever seen some log messages that are confusing or not helpful?"*).

*Part 3:* We ask the interviewees about their expectations on log messages with good readability and what practices can practitioners do to improve the readability of log messages.

At the end of each interview, we thank the interviewee and verify there is no sensitive information mentioned in the process of the interview.

**Interviewees.** We invite full-time employees working in software engineering related roles (e.g., software engineer, software architect, and test engineer) from 11 companies worldwide that are leading in their domains as our interviewees. The domain of those companies includes software development, internet services, telecommunications, electronics, investment management, and digital currency management. In total, 17 interviewees accepted our interview invitations. Their years of experience in software development and maintenance is 7.5 on average, ranging from 4 to 18 years. The interviewees are denoted as *I-1* to *I-17* when discussing their answers.

**Data Analysis.** After we complete all the interviews, the first author transcribes the interview record and performs open coding to generate an initial set of codes from the transcripts. The second author then verifies the codes and provides suggestions for improvement. We generate a total of 792 coded sentences from the transcripts. We further remove the codes that are not directly related to the readability of log messages (e.g., some interviewees mention that the timestamp of logs should have a consistent time zone setting, which is more related to the configuration of logging framework compared to the composition of log messages). A total of 161 coded sentences are removed in this step, with a total of 631 codes are preserved for further analysis. We then perform open card sorting [41] on the generated codes to analyze the thematic similarity. Specifically, the first two authors independently analyze the codes and sort the generated codes into potential themes that indicate the expected practices on the readability of log messages. We use Cohen's Kappa [42] to measure the agreement between the two authors. Overall, we have a Cohen's Kappa value of 0.76, which indicates a substantial agreement. The first two authors then discuss the disagreements until a consensus is reached. Eventually, we derive three aspects that are related to the readability of the log message, including **Structure**, **Information**, and **Wording**. Each aspect corresponds to several specific improvement practices that can be used to improve the readability. Some of the improvement practices are *corrective practices*, which are to improve the inadequacy of readability in log messages. Some are *enhancing practices*, developers may decide whether to apply them or not based on the situations and needs. We discuss each aspect and the corresponding practices in the results of RQ1 (Section IV).

## B. Stage 2: Manual Investigation

In this stage, we manually investigate the readability of log messages in real-world open source systems based on the aspects derived from the interviews. Specifically, we use the data set of logging statements provided by a prior study [22] to manually investigate the readability of log messages. Table I shows an overview of the data set. For each system, we randomly sample a set of logging statements to conduct the manual investigation based on 95% confidence level and 5% confidence interval [43]. In total, we randomly sample 2,702 logging statements from the nine systems. The sample size of each system varies from 201 in Wicket to 360 in HBase.

---

[1]The interview guideline can be found in our replication package [40].

**Manual Investigation Process.** The first two authors of this paper examine the sampled logging statements (i.e., 2,702 logging statements in total) with their surrounding code snippets. For each sampled logging statement, the two authors independently label whether the readability of its log message is adequate for each of the three aspects (i.e., *Structure*, *Information*, and *Wording*). When the labeling is finished, the first two authors then compare their results and discuss each disagreement until reaching a consensus. We have a Cohen's Kappa [42] value of 0.83 in this process, which indicates a substantial agreement.

*C. Stage 3: Survey*

To quantify the findings derived from our interviews and verify the manually investigated results, we conduct an online questionnaire survey with a larger number of participants.

**Survey Design.** The survey has five parts: Part 1 to Part 4 include multiple-choice questions, and Part 5 includes an open-ended question[2].

*Part 1:* We ask some background information related to the role and experience of the participants.

*Part 2:* We ask the participants for their perspective on the three aspects of readability derived from our prior interviews. We illustrate each aspect by providing two real-world examples randomly selected from our manual investigation results (i.e., *Stage 2*) which violate and comply the readability in the corresponding aspect, respectively. The participants then choose their consideration on the importance of the aspect to the readability of log messages from "Very important", "Important", "Neutral", "Unimportant", and "Very unimportant". At the end of this part, we further ask the participant for their overall perspective on the three aspects.

*Part 3:* We ask the participants for their perspective on the practices that can improve the readability of log messages from the corresponding aspect. We illustrate each practice by using a set of examples randomly selected from our manual investigation results (i.e., *Stage 2*). We then provide a statement indicating the effectiveness of each practice and ask the participant to choose their agreement level on the statement following a 5-point Likert scale (i.e., "Strongly agree", "Agree", "Neutral", "Disagree", and "Strongly disagree").

*Part 4:* We randomly select seven logging statements from our manual investigation results (i.e., *Stage 2* in Section III-B) and ask the participants to examine their readability. The participant can choose whether the log message of each logging statement is good or bad in terms of each of the three derived readability aspects. The main purpose of this part is to verify our manually labelled log messages in *Stage 2*. Note that the randomly selected logging statements for each participant are unique and do not have an overlap with other participants.

[2]Complete design of the survey is included in our replication package [40].

*Part 5:* We ask if the participants have other comments or ideas regarding the readability of log messages.

Due to the randomness of Part 4, we use online document platform (e.g., Google Doc) to design the surveys. Specifically, we prepare a series of survey documents in which Part 4 has unique logging statements and other parts have identical questions. Each participant has a unique link to the survey where the participants can directly write their answers.

In each multiple-choice question, we also provide an additional option "Not sure" if the participant cannot understand the question or does not have a clear answer. We also provide a comment field for each question where the participants are free to leave their comments related to the question.

We conduct a pilot survey with a small number of practitioners first to collect their feedback on the overall design of our survey. We made minor modifications to adjust the format of our survey and refine the description of questions based on their feedback and then have a final version of the survey. We then distribute our final version of the survey to the participants. Note that we exclude the responses collected from the participants in the pilot survey when we analyze and present the survey results.

**Participants.** We contact professionals in leading IT companies worldwide from our networks and ask their help to disseminate our survey to their colleagues. In total, we send out 80 surveys and receive 56 responses from the participants. Their years of experiences vary from 1 to 17 years, with an average of 5.4 years. The top two role of the participants are software engineer (38 participants) and test engineer (9 participants).

**Data Analysis.** We discard all the answers that select "Not sure". For the answers in Part 2 and Part 3, we report the percentage of each selected option. For the answers in Part 4, we analyze the results labelled by the participants and compare with our manual study results in *Stage 2* to examine the agreement level. We discuss the comments and feedback that we receive from the participants in Section IV.

*D. Stage 4: Automatic Classification*

As the first step to help improve the quality of log messages, in this stage, we seek to explore the potential of automatically classifying the readability of log messages. Specifically, for each aspect of readability (i.e., *Structure*, *Information*, and *Wording*), we classify whether a log message's readability is adequate or not in such aspect.

**Data Preparation.** We use the manually labelled log messages in *Stage 2* to train the models for automatic classification. For each log message, we tokenize it by space and attach its verbosity level (e.g., *info* or *error*) as the input feature, and use the labelled results as the target to predict.

Note that there are two steps for verifying the manually labelled log messages: 1) The first two authors independently label the log messages and discuss any disagreement until a consensus is reached. The Cohen's Kappa value of this process is 0.83, which is a substantial agreement; 2) In our survey discussed in *Stage 3*, we also ask the participants to label seven randomly sampled log messages. We receive 392 labelled log messages from the 56 participants. We further exclude the results of log messages with answers that are "Not sure". We then have 366 available log messages labelled by the survey participants, which is larger than the statistically significant sample size of 337, computed from the 2,702 log messages based on a 95% confidence level and a 5% confidence interval [43]. In this process, we find that a large number of the log messages labelled by the participants (81%) are exactly consistent with ours (i.e., the labels of all the three aspects are the same), which indicates that the results of manual investigation have high agreement with the survey participants.

**Classification Process.** Deep learning and machine learning approaches are widely used in the tasks of Software Engineering [44], [45], [46], [47]. We use one deep learning and four machine learning approaches to explore the potential of automatic classification. We follow a prior study on classifying good commit messages [48] for the selection of approaches and the hyper-parameter values. For deep learning, we use Bi-LSTM [49]; for machine learning, we use Logistic Regression [50], Decision Tree [51], Random Forest [52], and SVM [53]. We use Keras [54] and Scikit-learn [55] to implement the deep learning approach and machine learning approaches, respectively. We also use a state-of-the-art oversampling technique on the training data, namely ADASYN [56], to mitigate the potential impact of imbalanced data. For the vectorization of input features, we use Skip-gram from Word2vec [57] to train the word embeddings and transform the input features into numeric vectors. We then use each approach to train the models and evaluate their performance.

## IV. RESULTS

In this section, we discuss the results of each RQ.

### A. RQ1: What are Practitioners' Expectations on the Readability of Log Messages and How to Improve It?

In this RQ, we discuss the three aspects that are related to the readability of the log message derived from our interviews with practitioners, including *Structure*, *Information*, and *Wording*. For each aspect, we discuss: 1) Real-world example log messages that comply and violate the corresponding aspect, respectively; 2) Discussion of the interview and survey results; 3) Practices that can improve the readability. Some of the practices are *"corrective practices"*, which are practices to improve the inadequacy of readability in log messages. Some of the practices are *"enhancing practices"*, where developers can decide whether to apply them or not based on the situations and needs.

*1) Aspect 1 - Structure:* Format and organization of words and variables that a log message presents its information.

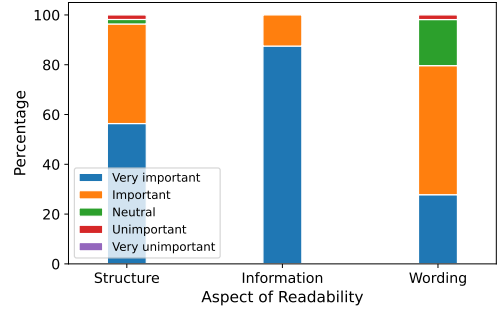

Fig. 2. Survey participants' rating for the importance of the three aspects.

**Example.** Below, we discuss two examples that violate and comply the aspect of *Structure*, respectively.

```
//Example 1 - VIOLATE the aspect of Structure
logger.debug("Bootstrap variables: {} {} {} {}",
    DatabaseDescriptor.isAutoBootstrap(),
    SystemKeyspace.bootstrapInProgress(),
    SystemKeyspace.bootstrapComplete(),
    DatabaseDescriptor.getSeeds());

//Example 2 - COMPLY the aspect of Structure
logger.debug("[repair #{}] Repair completed between {}
    and {} on {}", getId(), nodes.endpoint1,
    nodes.endpoint2, desc.columnFamily);
```

In Example 1, four variables are directly presented one by one. It might be difficult to distinguish the meaning of each variable in the generated logs. While in Example 2, the variables are presented together with descriptive words, which makes the meaning of variables easier to comprehend.

**Discussion.** Below, we discuss the interview results and survey results related to the aspect of *Structure*, respectively.

*Interview Results.* Among the 17 participants, 9 participants mention that Structure is important to the readability of log messages. For example, interviewee *I-8* expects that the log message should be *"well structured so it is easy to read by human"*. Interviewee *I-3* also mentions that:

*"Log message with good readability should have clear structure. For example, log messages that clearly separates variables could be easier to read. Don't present variables closely that are hard to judge boundaries."*

*Survey Results.* In our survey, we ask the participants for their perspective on the importance of each aspect. Figure 2 presents the percentage of each rate of importance given by the survey participants. We exclude three answers which are "Not sure" in all the three aspects and compute the percentage based on the remaining answers. Overall, more than half of the participants (55.3%) consider that Structure is "Very important" to the readability of log messages, and 39.3% of the participants consider it is "Important". Some survey participants also comment their perspective on this aspect. For example, one participant mentions that:

*"The aspect of Structure affects how the message is formulated. Better formulated log messages are always easier to read than unformulated ones".*

**Improvement Practices.** We derive three practices related to the aspect of *Structure*, including one corrective practice (i.e., practices to improve the inadequacy of readability in log messages) and two enhancing practices (i.e., practices that developers can decide whether to apply them or not based on the situations and needs). Below, we discuss each practice with corresponding examples.

*SP1 (Corrective): Have clear boundaries and distinctions among items.*

Different items in the log messages (e.g., variables) should have clear boundaries and descriptions to be easily distinguished. As shown in the example below, similar to the examples that we discussed in the aspect of *Structure*, the four variables in Example 1 are presented one by one which might be difficult to understand the meaning of each variable. Example 2 shows the log message that adopts this corrective practice, where each variable is added with a description of its meaning. In our interviews, 5 out of the 17 participants mention that this practice can be used to improve the readability of log message.

```
//Example 1 - WITHOUT SP1
LOG.debug("Reading from {} {} {} {}",
    tableDesc.getTableName(),
    region.getRegionNameAsString(),
    column.getNameAsString(),
    Bytes.toStringBinary(startKey));

//Example 2 - WITH SP1
LOG.debug("Reading from table: {}, region: {}, column:
    {}, key: {}", tableDesc.getTableName(),
    region.getRegionNameAsString(),
    column.getNameAsString(),
    Bytes.toStringBinary(startKey));
```

*SP2 (Enhancing): Use an easy-to-parse structure if needed and possible.*

Five interviewees mention that developers could consider formatting the log message that is easy to be automatically parsed by scripts for further analysis. For example, the code snippet shown below uses a comma (",") to separate each part. The ideal situation is to have log messages that are both human-readable and machine-readable.

```
//Example - WITH SP2
logger.info("Summary of the change, term: {}, version:
    {}, reason: {}", newClusterState.term(),
    newClusterState.version(), task.source);
```

*SP3 (Enhancing): Use parameterized logging to present the variables.*

Two interviewees mention that the log message in the logging statement with parameterized logging is easier to revisit and revise. Moreover, though not related to readability, parameterized logging costs less computing resource compared to simply concatenating the strings (according to the documentation of Log4j2 [14]).

```
//Example 1 - WITHOUT SP3
LOG.error("Exception when formatting: '" + dateStr + "'
    from: '" + fromFormat + "' to: '" + toFormat + "'",
    e);

//Example 2 - WITH SP3
logger.info("Exception when formatting: '{}' from '{}'
    to '{}' ", dateStr, fromFormat, toFormat, e);
```

*2) Aspect 2 - Information:* Semantic information conveyed by the log message to record system execution behaviors.

**Example.** Below, we discuss two examples that violate and comply the aspect of *Information*, respectively.

```
//Example 1 - VIOLATE the aspect of Information
LOG.info("Started.");

//Example 2 - COMPLY the aspect of Information
LOG.info("Quota support disabled, not starting space
    quota manager.");
```

In Example 1, the log message is "Started", but it is unclear what was started. In Example 2, the log message records the reason and consequence of a system event: due to the disabled quota support, the space quota manager is not starting.

**Discussion.** Below, we discuss the interview results and survey results related to the aspect of *Information*, respectively.

*Interview Results.* All of our 17 interviewees consider that the actual information that a log message conveys is important to its readability. For example, interviewee *I-9* mentions that:

*"The context of the log is important. When diagnosing the log, I would like to know how it happened. Like is it caused by an incorrect path or failed creation of files. It's also useful to know what is the consequence. Such as the consequence of the missing file. Will the system use the default configuration file or handle it with a different procedure."*

*Survey Results.* As shown in Figure 2, most of the participants (87.5%) consider that Information is "Very important" to the readability of log messages, and the remaining participants consider it is "Important". The results show that the participants highly acknowledge the importance of *Information* to the readability of log messages. For example, a survey participant comments that:

*"With more accurate information, the information aspect helps readers to better understand the message communicated by developers"*.

**Improvement Practices.** From our interviews, we derive three practices related to the aspect of *Information*, including two corrective practices and one enhancing practice. Below, we discuss each practice with corresponding examples.

*IP1 (Corrective): Provide proper context for the run-time behaviors.*

As shown in the examples below, the system execution behavior is the interruption of a current thread. In Example 1, the log message is just "Interrupted", while it's unclear what is interrupted. In Example 2, some context information of the execution behavior (i.e., the current thread) is added to the log message. It would be even better to include the thread ID if available.

```
//Example 1 - WITHOUT IP1
Thread.currentThread().interrupt();
LOG.info("Interrupted");

//Example 2 - WITH IP1
Thread.currentThread().interrupt();
LOG.info("The current thread is interrupted"); //(also
    add thread ID if available)
```

In our interviews, the participants suggest some context information that can be added into the log messages. We summarize the information into the following categories:

- Intention of this log message (clearly show whether it needs instant attention or not).
- Traceable information (e.g., thread and application ID).
- Clear "main character" of what happened *from* or what happened *to*.
- What is happening at the time.
- What is the consequence of this event.
- Possible reason of an unexpected event.

Note that it is not necessary to always include all of the context information every time, but our interviewees mention that the log messages should at least provide useful information and important events should include as sufficient context information as possible.

*IP2 (Corrective): Write a self-explanatory log message that is independent of other log messages.*

Six interviewees mention that the log message should be self-explanatory that does not depend on other log messages. As shown in Example 1 below, the log message in *debug* level is "Full exception". However, these two info and debug logs may not always be generated closely together (i.e., other logs may be generated in between). If there are other logs that appear before the debug log, it can be confusing to only see "Full exception" without the prior message. Hence, in Example 2, complete information is added to the *debug* level log to make it self-explanatory and avoid potential confusion.

```
//Example 1 - WITHOUT IP2
} catch (final AmazonClientException e) {
logger.info("Exception while retrieving instance list
    from AWS API: {}", e.getMessage());
logger.debug("Full exception:", e); //depending on the
    prior info log

//Example 2 - WITH IP2
} catch (final AmazonClientException e) {
logger.info("Exception while retrieving instance list
    from AWS API: {}", e.getMessage());
logger.debug("Exception while retrieving instance list
    from AWS API, full exception: ", e); //provides
    complete information that does not depend on other
    log messages
```

*IP3 (Enhancing): Minimize noise, emphasize the key information.*

Four interviewees mention that they want to concisely see the key information without too much noise. As shown in Example 1, the log message gives the instruction first and only mentions the error code and error message at the end. In Example 2, we simplify the log message to emphasize the error code and error message. Developers could consider adding another log or use another way to provide additional instructions if needed.

```
//Example 1 - WITHOUT IP3
LOG.warn("An HTTP error response in WebSocket
    communication would not be processed by the
    browser! If you need to send the error code and
    message to the client then configure custom
    WebSocketResponse via
    WebSocketSettings#newWebSocketResponse() factory
    method and override #sendError() method to write
    them in an appropriate format for your application.
    The ignored error code is '{}' and the message:
    '{}'.", sc, msg);

//Example2 - WITH IP3
LOG.warn("An HTTP error response in WebSocket
    communication would not be processed by the
    browser. Ignored error code: '{}', message: '{}'.
    ", sc, msg);
/*mention the key information first, can add another
    log, or use another way to write the additional
    instruction if it's necessary*/
```

*3) Aspect 3 - Wording:* Lexical usage of words and punctuation in the log message.

**Example.** Below, we discuss two examples that are violate and comply the aspect of *Wording*, respectively.

```
//Example 1 - VIOLATE the aspect of Wording
LOG.info("Added to offline, CURRENTLY NEVER CLEARED!!!");

//Example 2 - COMPLY the aspect of Wording
LOG.info("No family specified, will execute for all
    families.");
```

In Example 1, the log message uses an emotional wording (e.g., many exclamation marks and capitalization) to present a normal event. This may attract unnecessary attentions and confuses developers. While in Example 2, the log message uses standard wording to record an *info* level event.

**Discussion.** Below, we discuss the interview results and survey results related to the aspect of *Wording*, respectively.

*Interview Results.* Among the 17 participants, 7 participants mention that Wording is important to the readability of log messages. Some interviewees describe the scenarios where the wording affects the readability of log messages. For example, interviewees *I-1* and *I-13* mention that:

*"Similar to writing source code, we should have consistent naming conventions for the words of log messages too. Otherwise it might be confusing to the users".*

*"I've read some logs that have weird names included, hard to understand their meaning. Like are they identifiers or the abbreviations of anything".*

*Survey Results.* As shown in Figure 2, 50.0% of the participants consider that Wording is "Important" to the readability of log messages, and 26.8% of the participants consider it's "Very important". There are also 17.9% of the participants consider the importance of Wording is "Neutral". The survey results show that participants generally acknowledge the importance of *Wording*, but the priority is lower than *Information* and *Structure*. Some participants also provide comments to this aspect, for example:

*"Wording is important, but to a certain extend. Like tiny lexical mistakes can be acceptable."*

*"If the log message uses very emotional wording, I will obviously pay more attention to it and unhappy to see if it's just a trivial event".*

**Improvement Practices.** We derive five practices related to the aspect of *Wording*, including three corrective practices and two enhancing practices. Below, we discuss each practice with corresponding examples.

*WP1 (Corrective): Use standard English words (e.g., avoid typos and incomplete words).*

Four interviewees mention that we should avoid typos and incomplete words when writing the log messages. As shown in the example below, the word "preform" is a typo and should be "perform".

```
//Example - WITHOUT WP1
LOG.debug("Failed to preform reroute after cluster
    settings were updated."); //"preform" is a typo and
    should be "perform"
```

*WP2 (Corrective): Follow the convention of written language (e.g., correct grammar and not too oral).*

Three interviewees mention that log messages are better to follow the convention of written language. As shown in the example below, we do not "exists" is incorrect and should be "exist".

```
//Example - WITHOUT WP2
LOG.debug("Pinging a master {} but we do not exists on
    it, act as if its master failure"); //we do not
    "exists" should be "exist"
```

*WP3 (Corrective): Try to use impartial and neutral wording (e.g., avoid being too emotional or abusing capitalization).*

Three interviewees mention that the emotion of log messages should try to be neutral and objective. The examples shown below are both log messages with improper emotional wording. In Example 1, the log message is informal and uses many exclamation marks, which does not help with understanding the log message. Example 2 abuses the capitalization for a non-critical system event (i.e., *info* level).

```
//Example 1 - WITHOUT WP3
LOG.error("!!!!!Uh-oh, didn't find any action
    handlers!!!!!");

//Example 2 - WITHOUT WP3
LOG.info("Added to offline, CURRENTLY NEVER CLEARED!!!");
```

*WP4 (Enhancing): Be careful on using Abbreviations and Acronyms.*

Four interviewees mention that proper usage of abbreviations and acronyms is important. Developers should ensure that the users can understand the meaning of abbreviations and acronyms before writing them into the log message. As shown in the example, the abbreviation "TGT" is not a well-known word. Probably only users with corresponding domain knowledge can understand the meaning.

```
//Example
LOG.warn("No TGT found: will try again at {}");
```

*WP5 (Enhancing): Consistent on the wording of domain-specific terms.*

Six interviewees mention that the use of domain-specific terms should be consistent, otherwise it might be confusing for the users to understand their meaning. As shown in the example, "Incident ID" and "IncID" refer to the same thing. If possible, developers should consider keeping a consistent

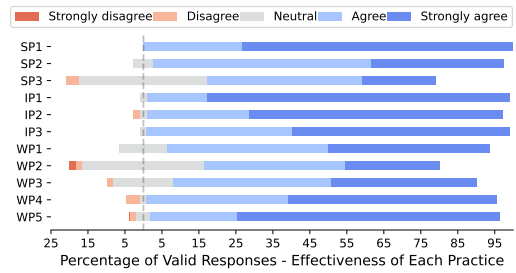

Fig. 3. Survey participants' rating for each improvement practice.

convention on the wording of domain terms to mitigate potential confusion.

```
//Example - WITHOUT WP5
LOG.info("Incident ID {}: a new incident is reported.",
    incID);
...
LOG.info("IncID {}: the incident is closed.", incID);
```

*4) Overall Perspectives on the Aspects and Improvement Practices*: In our survey, we also ask the participants for their overall perspectives on the three aspects above. Particularly, we ask for their perspectives on if these three aspects can reflect the readability of log messages. Participants can choose from "Very positive", "Positive", "Neutral", "Negative", "Very negative", and "Not sure". Overall, 51.8% of the participants' responses are "Very positive", and the remaining responses are "Positive". The results show that our survey participants acknowledge that the three aspects we derive based on the interviews can reflect the readability of log messages.

We also ask the survey participants for their agreement on the effectiveness for each improvement practice. For example in *Information Practice 1 (IP1)*, we provide a statement: *"This practice can improve the readability of log messages from the aspect of Information"*. Participants can choose their agreement level based on a 5-point Likert scale (i.e., "Strongly agree", "Agree", "Neutral", "Disagree", "Strongly disagree"), and an additional option of "Not sure". We exclude the answers that are "Not sure" (1.3% of the total answers) and present the distribution of results in Figure 3.

We find that for all the improvement practices, most of the responses have positive ratings (i.e., "Strongly agree" and "Agree"). In total, there are 1.5% of the responses that are negative (i.e., " Strongly Disagree" or "Disagree") regarding the specific improvement practices. For example, the participant who rates "Strongly Disagree" for *WP2* mentions *"I think people can still understand the message even if such mistakes are not corrected"*. Among the improvement practices for each aspect, *Information Practices* have the highest percentage of positive ratings, with an average of 97.7% for the effectiveness. For example, one participant who rates "Strongly Agree" for *IP2* mentions *"Context is important, "proper" context information is also very important, not too much and not too little."*.

TABLE II
PERCENTAGE (%) OF LOG MESSAGES IN EACH SYSTEM THAT HAVE
ADEQUATE READABILITY FOR ALL THE THREE ASPECTS, OR INADEQUATE
IN EACH OF THE ASPECT.

| Data set | Adequate | Inadequate | | |
|---|---|---|---|---|
| | | Structure | Information | Wording |
| Cassandra | 60.1 | 16.7 | 23.2 | 26.2 |
| Elasticsearch | 46.9 | 14.5 | 22.8 | 49.9 |
| Flink | 76.3 | 12.3 | 17.7 | 14.7 |
| HBase | 55.6 | 25.0 | 24.7 | 30.0 |
| JMeter | 52.0 | 27.0 | 30.7 | 36.4 |
| Kafka | 67.5 | 23.7 | 15.9 | 11.0 |
| Karaf | 75.7 | 12.0 | 16.7 | 21.1 |
| Wicket | 70.1 | 12.9 | 17.9 | 24.9 |
| Zookeeper | 60.0 | 15.6 | 23.1 | 34.6 |
| *Overall* | 61.9 | 18.2 | 21.7 | 28.1 |

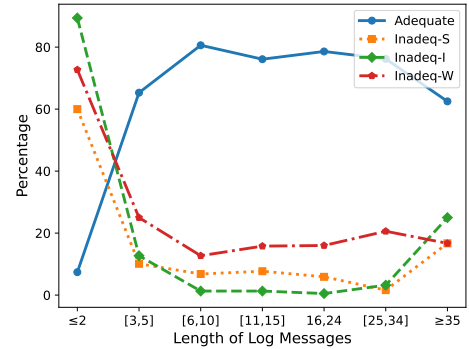

Fig. 4. Percentage of log messages with adequate or inadequate readability for different lengths. Length refers to the number of words of a log message.

> We derive three aspects that are related to the readability of log messages and several practices to improve each aspect. Among the three aspects, *Information* is considered as the most important aspect.

### B. RQ2: How is the Readability of Log Messages in large-scale Open Source Software Systems?

In this RQ, we present the results of our manual investigation on the readability of 2,702 logging statements sampled from nine large-scale open source systems, following the process discussed in *Stage 2* of Section III. We analyze the manual investigation results and present the results for: 1) Readability for log messages in different systems; 2) Readability for different lengths of log messages.

**Readability for Log Messages in Different Systems.** Table II presents the percentage of log messages in each data set that have adequate readability for all the three aspects (i.e., the column of *Adequate*), or inadequate in each of the aspect (i.e., *Structure*, *Information*, and *Wording* under the column of *Inadequate*). The row of *Overall* shows the overall percentage computed from all the data combined together. We find that the percentage of log messages with adequate readability varies in different systems, from 46.9% in Elasticsearch to 76.3% in Flink. We also find that the distribution of aspects is different for log messages with inadequate readability among the systems. For example, 49.9% of the log messages in Elasticsearch have inadequate readability in the aspect of Wording, while for Structure and Information the percentages are 14.5% and 22.8%, respectively.

**Readability for Log Messages with Different Lengths.** Figure 4 presents the percentage of log messages with adequate or inadequate readability for different lengths. We compute the length of a log message based on its number of words. *Adequate* refers to the percentage of log messages that have adequate readability in all the three aspects, *Inadeq-S*, *Inadeq-I*, and *Inadeq-W* refer to log messages that have inadequate readability in the aspect of *Structure*, *Information*, and *Wording*, respectively. We find that when the log messages are very short (i.e., length≤2), only 7.4% of the log messages have adequate readability in all the three aspects, with a very high percentage of *Inadeq-I* (89.4%). In contrast, log messages whose length is within the range between 6 to 10 words have the highest percentage with adequate readability (80.6%). When the log messages have more than 10 words, we then find that the readability has a downward trend as the length increases. For example, we find that the percentage of *Adequate* drops from 76.2% (log messages with number of words between 25 and 34) to 62.5% when the log messages have more than 35 words. Overall, the results show that the length of a log message might be an indicator of its readability, especially when the length is very short.

Moreover, we also ask the interviewees for their expectations on the length of log messages in the interviews. In total, 6 out of the 17 interviewees expect that the log message should be neither too short nor too long. Four interviewees consider that the log message should not be too short and 2 interviewees consider it should not be too long. There are also 5 interviewees do not have a specific expectation on the length itself, but the log message should provide clear and useful information. We find that our results in this RQ confirm the expectations from the interviewees. Compared to extremely short or long, log messages with a proper length tend to be more readable and are preferred by the practitioners.

> We find that only 61.9% of the log messages in our studied data set have adequate readability in all the three aspects, meaning that readability of a large portion of the log messages (i.e., 38.1%) in these systems are inadequate.

### C. RQ3: Can We Automatically Classify the Readability of Log Messages?

We take a preliminary step to help developers improve log message by classifying whether a message has readability issue or not. In this RQ, we present the results of automatic classification for the readability of log messages. We use the 2,702 manually labelled log messages to train and test the models using each approach discussed in Section III. We then perform a stratified 10-fold cross validation to estimate the performance of each approach and report the average results. Specifically, we randomly split the data set into ten subsets, with stratified random sampling [58] to ensure the same distribution of readability for each subset. The validation has ten rounds in total. For each round of validation, we use one subset for testing, and the remaining subsets for training.

TABLE III
BALANCED ACCURACY (%) OF DIFFERENT APPROACHES ON
CLASSIFYING THE READABILITY FOR EACH ASPECT.

|  | Structure | Information | Wording | *Average* |
|---|---|---|---|---|
| **Bi-LSTM** | 82.5 | 88.1 | 75.7 | 82.1 |
| **Random Forest** | **87.2** | **92.8** | **79.0** | **86.3** |
| **Decision Tree** | 78.8 | 86.5 | 75.2 | 79.5 |
| **Logistic Regression** | 65.0 | 78.3 | 60.1 | 67.8 |
| **SVM** | 67.9 | 85.3 | 72.4 | 75.2 |

TABLE IV
PRECISION, RECALL, AND F1 SCORE (%) OF CLASSIFYING EACH ASPECT
OF READABILITY USING RANDOM FOREST.

|  | Metric | Structure | Information | Wording | *Average* |
|---|---|---|---|---|---|
| Adequate | **Precision** | 95.4 | 97.1 | 87.5 | 93.3 |
|  | **Recall** | 95.2 | 96.0 | 91.7 | 94.3 |
|  | **F1** | 95.3 | 96.6 | 89.5 | 93.8 |
| Inadequate | **Precision** | 78.8 | 86.4 | 75.9 | 80.4 |
|  | **Recall** | 79.2 | 89.6 | 66.4 | 78.4 |
|  | **F1** | 78.9 | 87.9 | 70.7 | 79.2 |

**Balanced Accuracy of Different Approaches.** We first examine the balanced accuracy of each approach on classifying the three aspects of readability. Balanced accuracy is widely used by prior studies to evaluate the performance of binary classification on imbalanced data [59], [18], [20]. Table III shows the balanced accuracy of different approaches on classifying the readability for each aspect, the highest result is marked in bold. We find that Random Forest achieves the best balanced accuracy in all the three aspects of readability, with an average of 86.3%. Other approaches achieve a balanced accuracy from 67.8% by Logistic Regression to 82.1% by Bi-LSTM. Overall, we find that deep learning and machine learning approaches can both achieve promising classification results. Among them, Random Forest achieves the best balanced accuracy (i.e., 86.3% on average).

**Precision, Recall, and F1 score of Random Forest.** We further examine the performance of Random Forest on classifying the adequacy and inadequacy of each aspect. Table IV shows the Precision, Recall, and F1 score of classifying each aspect of readability using Random Forest. When *Adequate* readability in each aspect is considered as the positive class, Random Forest achieves an average precision, recall, and F1 score of 93.3%, and 94.3%, and 93.8%, respectively. When *Inadequate* readability in each aspect is considered as the positive class, the average precision, recall, and F1 score are 80.4%, and 78.4%, and 79.2%, respectively. Overall, we find that Random Forest can effectively classify each aspect of the readability. Our findings shed light on the possibility of automatically classifying readable log messages, future studies may consider leverage other state-of-the-art techniques (e.g., pre-trained large language models) to explore the potential of a better performance.

> Deep learning and machine learning approaches can both achieve promising results in the classification. Our findings take a preliminary step on automatically classifying the readability of log messages.

## V. IMPLICATIONS

We discuss the implications of our study for practitioners and researchers, respectively.

**Implication for Practitioners.** Due to the lack of well-defined guidelines on writing the log message, it is a challenging task to write log messages with good readability that can clearly and sufficiently record system run-time behaviors. Moreover, it is also difficult to decide what are log messages with "good readability". In our study, we conduct a series of interviews with industrial practitioners and derive three aspects that are related to the readability of log messages (i.e., *Structure*, *Information*, and *Wording*). For each aspect, we also discuss several specific practices that may improve the readability in such aspect. Practitioners can consider to refer our findings to have a clearer comprehension of the readability when composing and revising the log messages.

We also explore the potential of automatically classifying the readability of log messages. We find that several widely used deep learning approaches and machine learning approaches (e.g., Bi-LSTM, Random Forest, and Decision Tree) are effective in such classifications. Practitioners can leverage the automated approach to examine the readability of log messages they compose and obtain a suggestion of whether any aspects of the readability can be improved.

**Implication for Researchers.** In RQ2, we find that 38.1% of the studied log messages in large-scale open source systems have inadequate readability, meaning that there is still a large portion of the log messages of which readability may need improvement. Some prior studies work on automatically generating log messages using existing source code and log messages [35], [25]. However, we observe that these studies directly use the log messages to train and evaluate the models without a verification on the quality of those log messages. As a consequence, log messages with poor readability may be generated and thus decrease the reliability of such approaches. Future studies may leverage the findings in our study to examine the readability of log messages and prompt automated generation using more well-verified log messages.

Recently, large language models (e.g., GPT-3 [60]) have made remarkable progress in the comprehension and generation of natural languages. In this paper, we use classic deep learning and machine learning models (e.g., Bi-LSTM [49] and Random Forest [52]) to uncover the potential of automatically classifying the readability of log messages and achieve promising results. Future studies may consider explore the improvement of such classification by leveraging the large language models and further assist in logging practices.

## VI. THREATS TO VALIDITY

**Internal Validity.** We manually label the readability of log messages for each aspect. To mitigate the potential subjectivity, the first two authors label the log messages independently, and discuss each disagreement until a consensus is reached. The Cohen's Kappa value in this process is 0.83, which shows a substantial agreement. Involving the original authors of the logging statements who have contextual knowledge of the project could further verify the results of labelling. However, identifying and contacting the original author of each logging statement can be extremely challenging. In our survey discussed in *Stage 3*, we ask the participants to label seven randomly sampled log messages to further verify the manually labeled results instead. We find that 81% of the log messages labelled by the participants are exactly consistent with ours. For the results that are not consistent, the two authors who label the log messages discuss such cases and resolve the disagreements. The randomness while splitting the training and testing data sets may affect the results. To mitigate such threats, we use a stratified 10-fold cross validation to evaluate the results of each approach.

**External Validity.** We derive three aspects of readability and the corresponding improvement practices from the interviews. The logging practices might vary in different companies and thus the interview results may be different. To mitigate such threat, we invite participants from a variety of large companies to participate in our study, and the domain of their companies range from software development to digital currency management. These participants represent a variety of roles and level of software development and maintenance expertise.

## VII. CONCLUSION

In this paper, we investigate practitioners' expectations on the readability of log messages by conducting a series of interviews with industrial practitioners. We derive three aspects related to the readability of log messages along with several improvement practices for each aspect. Our findings receive encouraging feedback from subsequent online questionnaire surveys. We also find that a considerable proportion of the log messages in large-scale open source systems have inadequate readability. Therefore, we further explore the potential of automatically classifying the readability of log messages and find that both deep learning and machine learning approaches can effectively perform such classifications. The findings of our study provide a systematic understanding of the readability of log messages and shed light for future studies on providing comprehensive and automated supports for practitioners' logging practices.

## ACKNOWLEDGEMENTS

This research was supported by the National Natural Science Foundation of China (No. 62141222).

## REFERENCES

[1] S. Ma, X. Zhang, D. Xu *et al.*, "Protracer: Towards practical provenance tracing by alternating between logging and tainting." in *NDSS*, vol. 2, 2016, p. 4.
[2] M. Nagappan, K. Wu, and M. A. Vouk, "Efficiently extracting operational profiles from execution logs using suffix arrays," in *ISSRE'09: Proceedings of the 20th IEEE International Conference on Software Reliability Engineering*, 2009, pp. 41–50.
[3] Z. Ding, Y. Tang, Y. Li, H. Li, and W. Shang, "On the temporal relations between logging and code," in *2023 IEEE/ACM 45th International Conference on Software Engineering (ICSE)*. IEEE, 2023, pp. 843–854.
[4] X. Zhang, Y. Xu, Q. Lin, B. Qiao, H. Zhang, Y. Dang, C. Xie, X. Yang, Q. Cheng, Z. Li, J. Chen, X. He, R. Yao, J.-G. Lou, M. Chintalapati, F. Shen, and D. Zhang, "Robust log-based anomaly detection on unstable log data," in *Proceedings of the 2019 27th ACM Joint Meeting on European Software Engineering Conference and Symposium on the Foundations of Software Engineering*, ser. ESEC/FSE 2019, 2019, p. 807–817.
[5] L. Yang, J. Chen, Z. Wang, W. Wang, J. Jiang, X. Dong, and W. Zhang, "Plelog: Semi-supervised log-based anomaly detection via probabilistic label estimation," in *43rd IEEE/ACM International Conference on Software Engineering: Companion Proceedings, ICSE Companion 2021, Madrid, Spain, May 25-28, 2021*, 2021, pp. 230–231.
[6] Z. Li, C. Luo, T. Chen, W. Shang, S. He, Q. Lin, and D. Zhang, "Did we miss something important? studying and exploring variable-aware log abstraction," in *2023 IEEE/ACM 45th International Conference on Software Engineering (ICSE)*, 2023, pp. 830–842.
[7] X. Zhou, X. Peng, T. Xie, J. Sun, C. Ji, D. Liu, Q. Xiang, and C. He, "Latent error prediction and fault localization for microservice applications by learning from system trace logs," in *Proceedings of the ACM Joint Meeting on European Software Engineering Conference and Symposium on the Foundations of Software Engineering, ESEC/SIGSOFT FSE 2019*, 2019, pp. 683–694.
[8] D. Yuan, J. Zheng, S. Park, Y. Zhou, and S. Savage, "Improving software diagnosability via log enhancement," in *ASPLOS '11: Proceedings of the 16th international conference on Architectural support for programming languages and operating systems*. ACM, 2011, pp. 3–14.
[9] A. R. Chen, T.-H. Chen, and S. Wang, "Pathidea: Improving information retrieval-based bug localization by re-constructing execution paths using logs," *IEEE Transactions on Software Engineering*, pp. 2905–2919, 2021.
[10] Q. Lin, H. Zhang, J.-G. Lou, Y. Zhang, and X. Chen, "Log clustering based problem identification for online service systems," in *2016 IEEE/ACM 38th International Conference on Software Engineering Companion (ICSE-C)*. IEEE, 2016, pp. 102–111.
[11] D. Schipper, M. F. Aniche, and A. van Deursen, "Tracing back log data to its log statement: from research to practice," in *Proceedings of the 16th International Conference on Mining Software Repositories, MSR 2019*, 2019, pp. 545–549.
[12] A. R. Chen, T.-H. Chen, and S. Wang, "Demystifying the challenges and benefits of analyzing user-reported logs in bug reports," *Empirical Software Engineering*, pp. 1–30, 2021.
[13] "Elasticsearch github page," https://github.com/elastic/elasticsearch, 2023, last accessed August 2023.
[14] Apache, "log4j2," https://logging.apache.org/log4j/2.x/manual/messages.html, 2023, last accessed May 2023.
[15] H. Li, W. Shang, B. Adams, M. Sayagh, and A. E. Hassan, "A qualitative study of the benefits and costs of logging from developers' perspectives," *IEEE Transactions on Software Engineering*, 2020.
[16] J. Zhu, P. He, Q. Fu, H. Zhang, M. R. Lyu, and D. Zhang, "Learning to log: Helping developers make informed logging decisions," in *Proceedings of the 37th International Conference on Software Engineering*, ser. ICSE '15, 2015, pp. 415–425.
[17] Q. Fu, J. Zhu, W. Hu, J.-G. Lou, R. Ding, Q. Lin, D. Zhang, and T. Xie, "Where do developers log? an empirical study on logging practices in industry," in *Proceedings of the 36th International Conference on Software Engineering*, ser. ICSE-SEIP '14, 2014, pp. 24–33.
[18] Z. Li, T. Chen, and W. Shang, "Where shall we log? studying and suggesting logging locations in code blocks," in *35th IEEE/ACM International Conference on Automated Software Engineering, ASE 2020*, 2020, pp. 361–372.
[19] J. Cândido, J. Haesen, M. Aniche, and A. van Deursen, "An exploratory study of log placement recommendation in an enterprise system," in *2021 IEEE/ACM 18th International Conference on Mining Software Repositories (MSR)*, 2021, pp. 143–154.
[20] H. Li, W. Shang, Y. Zou, and A. E. Hassan, "Towards just-in-time suggestions for log changes," *Empirical Software Engineering*, pp. 1831–1865, 2017.
[21] H. Li, W. Shang, and A. E. Hassan, "Which log level should developers choose for a new logging statement?" *Empirical Software Engineering*, vol. 22, no. 4, pp. 1684–1716, Aug 2017.
[22] Z. Li, H. Li, T.-H. P. Chen, and W. Shang, "Deeplv: Suggesting log levels using ordinal based neural networks," in *2021 IEEE/ACM 43rd International Conference on Software Engineering (ICSE)*. IEEE, 2021, pp. 1461–1472.

[23] J. Liu, J. Zeng, X. Wang, K. Ji, and Z. Liang, "Tell: log level suggestions via modeling multi-level code block information," in *ISSTA '22: 31st ACM SIGSOFT International Symposium on Software Testing and Analysis*, S. Ryu and Y. Smaragdakis, Eds., 2022, pp. 27–38.

[24] A. Mastropaolo, L. Pascarella, and G. Bavota, "Using deep learning to generate complete log statements," in *Proceedings of the 44th International Conference on Software Engineering*, ser. ICSE '22, 2022, p. 2279–2290.

[25] Z. Ding, H. Li, and W. Shang, "Logentext: Automatically generating logging texts using neural machine translation," in *2022 IEEE International Conference on Software Analysis, Evolution and Reengineering (SANER)*, 2022, pp. 349–360.

[26] D. Yuan, S. Park, and Y. Zhou, "Characterizing logging practices in open-source software," in *ICSE 2012: Proceedings of the 2012 International Conference on Software Engineering*, 2012, pp. 102–112.

[27] B. Chen and Z. M. (Jack) Jiang, "Characterizing logging practices in java-based open source software projects – a replication study in apache software foundation," *Empirical Software Engineering*, pp. 330–374, 2017.

[28] Y. Zeng, J. Chen, W. Shang, and T.-H. P. Chen, "Studying the characteristics of logging practices in mobile apps: a case study on f-droid," *Empirical Software Engineering*, pp. 1–41, 2019.

[29] K. Patel, J. Faccin, A. Hamou-Lhadj, and I. Nunes, "The sense of logging in the linux kernel," *Empirical Software Engineering*, pp. 1–47, 2022.

[30] H. Zhang, Y. Tang, M. Lamothe, H. Li, and W. Shang, "Studying logging practice in test code," *Empirical Software Engineering*, p. 83, 2022.

[31] S. He, P. He, Z. Chen, T. Yang, Y. Su, and M. R. Lyu, "A survey on automated log analysis for reliability engineering," *ACM computing surveys (CSUR)*, vol. 54, no. 6, pp. 1–37, 2021.

[32] Z. Li, "Towards providing automated supports to developers on writing logging statements," in *Proceedings of the ACM/IEEE 42nd International Conference on Software Engineering: Companion Proceedings*, 2020, pp. 198–201.

[33] Z. Li, T. P. Chen, J. Yang, and W. Shang, "DLFinder: characterizing and detecting duplicate logging code smells," in *Proceedings of the 41st International Conference on Software Engineering, ICSE 2019*, 2019, pp. 152–163.

[34] Z. Li, T.-H. Chen, J. Yang, and W. Shang, "Studying duplicate logging statements and their relationships with code clones," *IEEE Transactions on Software Engineering*, pp. 2476–2494, 2021.

[35] H. Pinjia, Z. Chen, S. He, and M. R. Lyu, "Characterizing the natural language descriptions in software logging statements," in *Proceedings of the 33rd IEEE international conference on Automated software engineering*, 2018, pp. 1–11.

[36] Z. Masood, R. Hoda, and K. Blincoe, "What drives and sustains self-assignment in agile teams," *IEEE Transactions on Software Engineering*, 2021.

[37] E. Kalliamvakou, C. Bird, T. Zimmermann, A. Begel, R. DeLine, and D. M. German, "What makes a great manager of software engineers?" *IEEE Transactions on Software Engineering*, pp. 87–106, 2017.

[38] E. Kalliamvakou, G. Gousios, K. Blincoe, L. Singer, D. M. German, and D. Damian, "The promises and perils of mining github," in *Proceedings of the 11th working conference on mining software repositories*, 2014, pp. 92–101.

[39] X. Hu, X. Xia, D. Lo, Z. Wan, Q. Chen, and T. Zimmermann, "Practitioners' expectations on automated code comment generation," in *Proceedings of the 44th International Conference on Software Engineering*, 2022, pp. 1693–1705.

[40] "Link to our replication package." https://github.com/ginolzh/ASE2023_Log_Message_Readability, last accessed May 2023.

[41] D. Spencer, *Card sorting: Designing usable categories*. Rosenfeld Media, 2009.

[42] M. L. McHugh, "Interrater reliability: the kappa statistic," *Biochemia Medica*, vol. 22, no. 3, pp. 276–282, 2012.

[43] S. Boslaugh and P. Watters, *Statistics in a Nutshell: A Desktop Quick Reference*, ser. In a Nutshell (O'Reilly). O'Reilly Media, 2008.

[44] C. Watson, N. Cooper, D. N. Palacio, K. Moran, and D. Poshyvanyk, "A systematic literature review on the use of deep learning in software engineering research," *ACM Transactions on Software Engineering and Methodology (TOSEM)*, pp. 1–58, 2022.

[45] A. Mastropaolo, N. Cooper, D. N. Palacio, S. Scalabrino, D. Poshyvanyk, R. Oliveto, and G. Bavota, "Using transfer learning for code-related tasks," *IEEE Transactions on Software Engineering*, 2022.

[46] T. Hoang, H. K. Dam, Y. Kamei, D. Lo, and N. Ubayashi, "Deepjit: an end-to-end deep learning framework for just-in-time defect prediction," in *2019 IEEE/ACM 16th International Conference on Mining Software Repositories (MSR)*, 2019, pp. 34–45.

[47] D. Di Nucci, F. Palomba, D. A. Tamburri, A. Serebrenik, and A. De Lucia, "Detecting code smells using machine learning techniques: are we there yet?" in *2018 ieee 25th international conference on software analysis, evolution and reengineering (saner)*, 2018, pp. 612–621.

[48] Y. Tian, Y. Zhang, K. Stol, L. Jiang, and H. Liu, "What makes a good commit message?" in *44th IEEE/ACM 44th International Conference on Software Engineering, ICSE 2022*, 2022, pp. 2389–2401.

[49] M. Schuster and K. K. Paliwal, "Bidirectional recurrent neural networks," *IEEE transactions on Signal Processing*, pp. 2673–2681, 1997.

[50] S. Menard, *Applied logistic regression analysis*. Sage, 2002, no. 106.

[51] J. R. Quinlan, "Induction of decision trees," *Machine learning*, pp. 81–106, 1986.

[52] L. Breiman, "Random forests," *Machine learning*, pp. 5–32, 2001.

[53] J. Platt *et al.*, "Probabilistic outputs for support vector machines and comparisons to regularized likelihood methods," *Advances in large margin classifiers*, pp. 61–74, 1999.

[54] "Keras: The python deep learning library," https://keras.io/, last accessed May 2023.

[55] "scikit-learn: Machine learning in python," https://scikit-learn.org, last accessed May 2023.

[56] H. He, Y. Bai, E. A. Garcia, and S. Li, "Adasyn: Adaptive synthetic sampling approach for imbalanced learning," in *2008 IEEE international joint conference on neural networks (IEEE world congress on computational intelligence)*. IEEE, 2008, pp. 1322–1328.

[57] "gensim Word2vec embeddings," https://radimrehurek.com/gensim/models/word2vec.html, last accessed May 2023.

[58] H. Pirzadeh, S. Shanian, A. Hamou-Lhadj, and A. Mehrabian, "The concept of stratified sampling of execution traces," in *The 19th IEEE International Conference on Program Comprehension, ICPC 2011*, 2011, pp. 225–226.

[59] J. Zhu, P. He, Q. Fu, H. Zhang, M. R. Lyu, and D. Zhang, "Learning to log: Helping developers make informed logging decisions," in *Proceedings of the 37th International Conference on Software Engineering*, ser. ICSE '15, 2015, pp. 415–425.

[60] T. Brown, B. Mann, N. Ryder, M. Subbiah, J. D. Kaplan, P. Dhariwal, A. Neelakantan, P. Shyam, G. Sastry, A. Askell *et al.*, "Language models are few-shot learners," *Advances in neural information processing systems*, 2020.